# Network controllability in transmodal cortex predicts psychosis spectrum symptoms

Linden Parkes PhD[1], Tyler M. Moore PhD[2,3], Monica E. Calkins PhD[2,3], Matthew Cieslak PhD[2,3,4], David R. Roalf PhD[2,3], Daniel H. Wolf MD PhD[2,3,4], Ruben C. Gur PhD[2,3,5,6], Raquel E. Gur MD PhD[2,3,5,6], Theodore D. Satterthwaite MD[2,3,4], & Danielle S. Bassett PhD[†1,2,5,7,8,9]

[1]Department of Bioengineering, School of Engineering & Applied Science, University of Pennsylvania, Philadelphia, PA, 19104 USA.
[2]Department of Psychiatry, Perelman School of Medicine, University of Pennsylvania, Philadelphia, PA 19104, USA.
[3]Lifespan Brain Institute, University of Pennsylvania & Children's Hospital of Philadelphia, Philadelphia, USA
[4]Center for Biomedical Image Computing and Analytics, Perelman School of Medicine, University of Pennsylvania, Philadelphia, PA 19104 USA.
[5]Department of Neurology, Perelman School of Medicine, Philadelphia, PA 19104 USA.
[6]Department of Radiology, Perelman School of Medicine, Philadelphia, PA 19104 USA.
[7]Department of Electrical & Systems Engineering, School of Engineering & Applied Science, University of Pennsylvania, Philadelphia, PA, 19104 USA.
[8]Department of Physics & Astronomy, College of Arts & Sciences, University of Pennsylvania, Philadelphia, PA, 19104 USA.
[9]Santa Fe Institute, Santa Fe, NM 87501 USA
[†]To whom correspondence should be addressed: dsb@seas.upenn.edu

Corresponding author: Danielle S. Bassett, dsb@seas.upenn.edu, Suite 240 Skirkanich Hall, 210 Sth 33rd St, Philadelphia, PA 19104-6321, USA

**ABSTRACT**

**Background:** The psychosis spectrum is associated with structural dysconnectivity concentrated in transmodal association cortex. However, understanding of this pathophysiology has been limited by an exclusive focus on the direct connections to a region. Using Network Control Theory, we measured variation in both direct and indirect structural connections to a region to gain new insights into the pathophysiology of the psychosis spectrum.

**Methods:** We used psychosis symptom data and structural connectivity in 1,068 youths aged 8 to 22 years from the Philadelphia Neurodevelopmental Cohort. Applying a Network Control Theory metric called average controllability, we estimated each brain region's capacity to leverage its direct and indirect structural connections to control linear brain dynamics. Next, using non-linear regression, we determined the accuracy with which average controllability could predict negative and positive psychosis spectrum symptoms in out-of-sample testing. We also compared prediction performance for average controllability versus strength, which indexes only direct connections to a region. Finally, we assessed how the prediction performance for psychosis spectrum symptoms varied over the functional hierarchy spanning unimodal to transmodal cortex.

**Results:** Average controllability outperformed strength at predicting positive psychosis spectrum symptoms, demonstrating that indexing indirect structural connections to a region improved prediction performance. Critically, improved prediction was concentrated in association cortex for average controllability, whereas prediction performance for strength was uniform across the cortex, suggesting that indexing indirect connections is crucial in association cortex.

**Conclusions:** Examining inter-individual variation in direct and indirect structural connections to association cortex is crucial for accurate prediction of positive psychosis spectrum symptoms.

**INTRODUCTION**

The psychosis spectrum (PS) is broadly characterized by positive (e.g., hallucinations, delusions, disorganized thought) and negative (e.g., avolition, flattened affect, social withdrawal) psychosis symptoms (1). The PS follows a continuous distribution of severity, with absence of these symptoms at one end and disorders such as schizophrenia, which greatly impact daily functioning (2), at the other (3). Transition along the PS towards schizophrenia occurs predominantly during adolescence and young adulthood (2,4), and is thought to be underpinned by widespread structural dysconnectivity that emerges during this time (5–7). In this context, regional (dys)connectivity is typically characterized by examining the direct connections between a given region and the rest of the brain. However, mounting evidence demonstrates that any region's capacity to affect the activity of other brain regions is also strongly influenced by the presence and nature of indirect connections. Indeed, the brain contains circuits whose topology enables the dynamic spread of activity between regions that may not be directly connected (8–15). How these indirect aspects of structural connectivity – and their impact on the spread of activity and control of brain states – might relate to PS symptoms remains unclear, rendering our understanding of the neurobiology of the PS incomplete.

The capacity for spatially distributed brain regions to communicate and coordinate their activity is essential for normal cognitive and affective functions (11). Critical to this communication is the brain's underlying structural connectivity, which provides a scaffold along which activity in one region can spread to, and influence, another region. The structural connectivity profile of a brain region is often summarized regionally using graph theoretic metrics such as degree and strength (16,17); the former calculated by counting the number of binary direct connections to a region, and the latter by summing over the weights of those connections. These metrics have aided our understanding of the brain's structural organization (7,18,19). For example, adolescent development gives rise to regions with disproportionately high degree and strength, known as hubs (20–22). In the cerebral cortex, hubs are commonly found in transmodal association areas (21,23–25), where they are thought to integrate across functionally specialized and segregated subnetworks and enable complex higher-order human functions (23,26–28). In support of this integrative role, transmodal regions connect with far-reaching and cascading indirect pathways that exert regulatory control over unimodal regions (29–31). However, these indirect paths are not captured by the analysis of regional strength. Given that these indirect pathways converge to a greater extent on transmodal regions compared to unimodal regions (29), assessing variation in indirect connection pathways is likely critical to understanding the integrative role of association cortex.

The analysis of direct structural connections has also informed our understanding of how structural connectivity varies along the PS (32–36). Notably, dysconnectivity in transmodal cortex is a prominent feature in individuals on the PS, being commonly reported for schizophrenia (32,35–38), first-episode psychosis (34), individuals at-risk for psychosis (33), and psychosis-like experiences (39). Further, this dysconnectivity is reflected by disrupted integration in the brains of individuals with schizophrenia (23,35,40,41). However, these studies have been limited by a focus on examining variability in direct structural connections, potentially missing important symptom-related variation in regional connectivity profiles. Thus, examining the dysconnectivity of indirect pathways that stem from transmodal cortex may help elucidate the pathophysiology of the PS.

Here, we use Network Control Theory (NCT) (9) to examine whether individuals on the PS display alterations in the ability of indirect structural connections to influence the spread of activity and control brain states. NCT is a branch of physical and engineering sciences that treats a network as a dynamical system (9,42). The application of NCT has revolutionized both the understanding and design of complex networks in contexts as diverse as space and terrestrial exploration, and modeling of financial markets, airline networks, and fire-control systems. In each of these examples, the system is controlled through signals that originate at a control point and move through the network. In the brain, NCT models each region's activity as a time-dependent internal state that is predicted from a combination of three factors: (i) its previous state, (ii) whole-brain structural connectivity, and (iii) external inputs. After linearizing the system's dynamics, *average controllability* quantifies a region's capacity to distribute activity throughout the brain, beyond the bounds of its direct connections, in order to guide changes in brain state (9,42). Prior work has demonstrated that average controllability increases throughout development (43), supporting optimal executive function (44,45), is heritable in the prefrontal cortex (46), and is disrupted in bipolar disorder (47). Critically, while high strength is necessary for high average controllability (9), analysis of inter-individual differences has shown that strength and average controllability have unique variance over subjects and thus do not represent redundant summaries of regional structural connectivity (44). However, it remains unclear to what extent inter-individual variability in direct connections, indirect connections, or a combination of both, predict individual differences in PS symptoms.

Here, we sought to understand how direct and indirect connections differentially contribute to prediction of PS symptoms in a large sample of youth with a broad spectrum of psychiatric symptoms. We operationalized this goal by comparing the ability of strength and average controllability to predict positive and negative PS symptoms in out-of-sample testing (48). We tested three hypotheses. First, owing to its capacity to use both direct and indirect structural connections to control brain states, we hypothesized that average controllability, not strength, would best predict PS symptoms. Second, due to their far-reaching indirect connectivity profiles (9,10), we hypothesized that regions in transmodal cortex would be more sensitive to variations in indirect connectivity compared to regions in the unimodal sensorimotor cortex. Thus, we predicted that regional cross-subject correlations between strength and average controllability would be lower in transmodal cortex compared to unimodal cortex. Finally, reflecting this divergence in transmodal cortex, we expected that better predictive performance for average controllability, compared to strength, would be driven predominantly by regions in transmodal cortex.

**MATERIALS AND METHODS**

*Participants*

Participants included 1,601 individuals from the Philadelphia Neurodevelopmental Cohort (49), a community-based study of brain development in youths aged 8 to 22 years with a broad range of psychopathology (50,51). We studied a subset of 1,068 participants, including individuals who were medically healthy and passed stringent neuroimaging quality control (see Supplementary Methods).

*Dimensional measures of the psychosis spectrum*

In order to study inter-individual variation in the psychosis spectrum (PS), we used a model of psychopathology based on the *p*-factor hypothesis (48). Briefly, the *p*-factor hypothesis (52–58) posits that psychopathology symptoms cluster into latent dimensions including a general factor (known as *p* or 'overall psychopathology'), which underpins individuals' tendency to develop all forms of psychopathology, alongside multiple dimensions that describe specific types of psychopathology. As used in our previous work (48) and detailed in the Supplementary Material, we used confirmatory bifactor analysis to quantify six orthogonal dimensions of psychopathology, including an overall psychopathology factor common to all symptoms measured herein, and five specific factors: psychosis-positive, psychosis-negative, anxious-misery, externalizing behaviors, and fear (53) (also see Table S1 and Table S2). Here, we primarily studied the psychosis-positive and psychosis-negative dimensions, which represent the positive and negative domains of the PS, respectively (1). We also studied the overall psychopathology dimension. The joint examination of these three dimensions allowed us to examine the extent to which direct and indirect structural dysconnectivity reflected PS-specific or disorder-general biomarkers.

*Structural network estimation*

For details on image acquisition, quality control, and processing see Supplementary Methods. Briefly, for each participant, whole-brain deterministic fiber tracking was conducted using DSI Studio (59) and the number of streamlines intersecting region $i$ and region $j$ in a parcellation of N=200 regions (60) was used to weight the edges of an undirected adjacency matrix, $A$. Note, $A_{ij} = 0$ for $i = j$.

*Strength*

A simple summary of a region's direct structural connections to the rest of the brain is its weighted degree, or strength (16). Within each participant's adjacency matrix, $A$, we calculated strength $s$ for region $i$ as the sum of edge weights over all regions in the network,

$$s_i = \sum_{j=1}^{N} A_{ij.} \quad\quad\quad Eq.\ 1$$

Of note, here we examined strength instead of unweighted degree to ensure comparability with average controllability, which is estimated from the weighted, rather than the binarized, $A$ matrix.

*Network controllability*

Network Control Theory (NCT) provides a means to study how the brain's structural network supports, constrains, and controls temporal dynamics in brain activity. Details of the NCT tools have been extensively discussed in prior work (9,61,62). Here, we draw on an NCT metric known as average controllability (9,43). Below, we briefly describe the derivation of average controllability as well as the model of population activity that underpins it.

We define the activity state of the brain using a simplified noise-free linear discrete-time and time-invariant model of regional dynamics:

$$x(t+1) = Ax(t) + B_\kappa u_\kappa(t), \quad\quad\quad Eq.\ 2$$

where $x(t)$ is a vector of size $N \times 1$ that represents the state of the system at time $t$. Here, $N$ is the number of brain regions in the system, and therefore the state is the pattern of brain activity across the regions at a single point in time. Over time, $x(t)$ denotes the brain state trajectory, a temporal sequence of the aforementioned pattern of brain activity. The matrix $A$ denotes the normalized $N \times N$ structural adjacency matrix, which describes the structural connections between system nodes. We normalize each participant's $A$ matrix in the following manner:

$$A = \frac{A}{|\lambda(A)|_{max} + c}. \qquad \text{Eq. 3}$$

Here, $|\lambda(A)|_{max}$ is the largest eigenvalue of $A$. Additionally, we set $c = 1$ to ensure that the system is stable (see Supplementary Methods for more details).

The matrix $B_\kappa$ in *Eq. 2* is of size $N \times N$ and describes the brain regions $\kappa$ into which we inject inputs before assessing the response of the system. In this study, we calculated average controllability for each brain region separately; thus, $B_\kappa$ simplifies to an $N \times 1$ vector where the element corresponding to the node being controlled is set to 1 and all other elements are set to 0. Finally, $u_\kappa(t)$ is a vector of size $N \times 1$ that indicates the amount of input energy injected into the control nodes listed in $B_\kappa$ at each time point $t$.

*Average controllability*

Average controllability describes the ability of a network to spread the input energy injected into a control node throughout the system to affect changes in brain states (9). As in previous work, we use $\text{Trace}(W_{\kappa,T})$, where $W_{\kappa,T}$ is the controllability Gramian,

$$W_{\kappa,T} = \sum_{\tau=0}^{T-1} A^\tau B_\kappa B_\kappa^\top (A^\top)^\tau, \qquad \text{Eq. 4}$$

where ⊤ denotes the transpose operation, $\tau$ indicates the time step of the trajectory, and $T$ denotes the time horizon, which is set to infinity. Average controllability is computed for each node in $A$ separately.

*Machine learning prediction models*

The above procedures generated two $1,068 \times 200$ matrices ($X$) of regional structural connectivity features: strength ($X_s$) and average controllability ($X_a$; Figure 1A). To ensure normality, columns of these matrices, as well as the PS symptom dimensions, were normalized using an inverse normal transformation (63,64). Then, connectivity features were taken as multivariate input features to kernel ridge regression (KRR) with a radial basis function (65) to iteratively predict symptom dimensions ($y$) in a series of prediction models. These prediction models are explained in detail in the supplementary methods and are outlined briefly below.

First, for each ($X, y$) combination, we performed 100 repeats (66) of 10-fold cross-validation scored by root mean squared error (RMSE) that included leakage-resistant nuisance regression of age, sex, total brain volume, and in-scanner motion. We hereafter refer to this as our *primary prediction model* (Figure 1B). We complemented this model with a *secondary prediction model* that included hyper-parameter optimization via nested cross-validation instead of nuisance regression (Figure S1). For both our primary and secondary prediction models and for each symptom dimension ($y$), we compared prediction performance (RMSE distributions) across strength ($X_s$) and average controllability ($X_a$) using an exact test of differences (67).

Both our primary and secondary prediction models generated robust estimates of prediction performance that could be compared across ($X, y$) combinations but did not examine whether prediction performance was in itself significant. Thus, to test whether prediction performance was better than chance, we compared a point estimate of RMSE to the distribution of RMSE values obtained from permuted data, wherein $y$ was randomly shuffled 5,000 times. We refer to this model as our *null prediction model* (Figure S2).

Finally, we sought an approach to examining each region's contribution to prediction performance in our non-linear KRR. We trained a given ($X, y$) combination on a single cross-validation split, stratified on $y$. However, instead of training on all columns of $X$, we trained on non-overlapping subsets of five columns sampled along the principal cortical gradient of functional connectivity (26). The principal cortical gradient separates the transmodal cortex from unimodal cortex in a continuous fashion, which allowed us to characterize gradual changes in prediction performance for strength ($X_s$) and average controllability ($X_a$) as a function of the cortical hierarchy. We derived the cortical gradient in our own data (see Supplementary Methods). We refer to this third, and final, model as our *binned-regions prediction model* (Figure 1C).

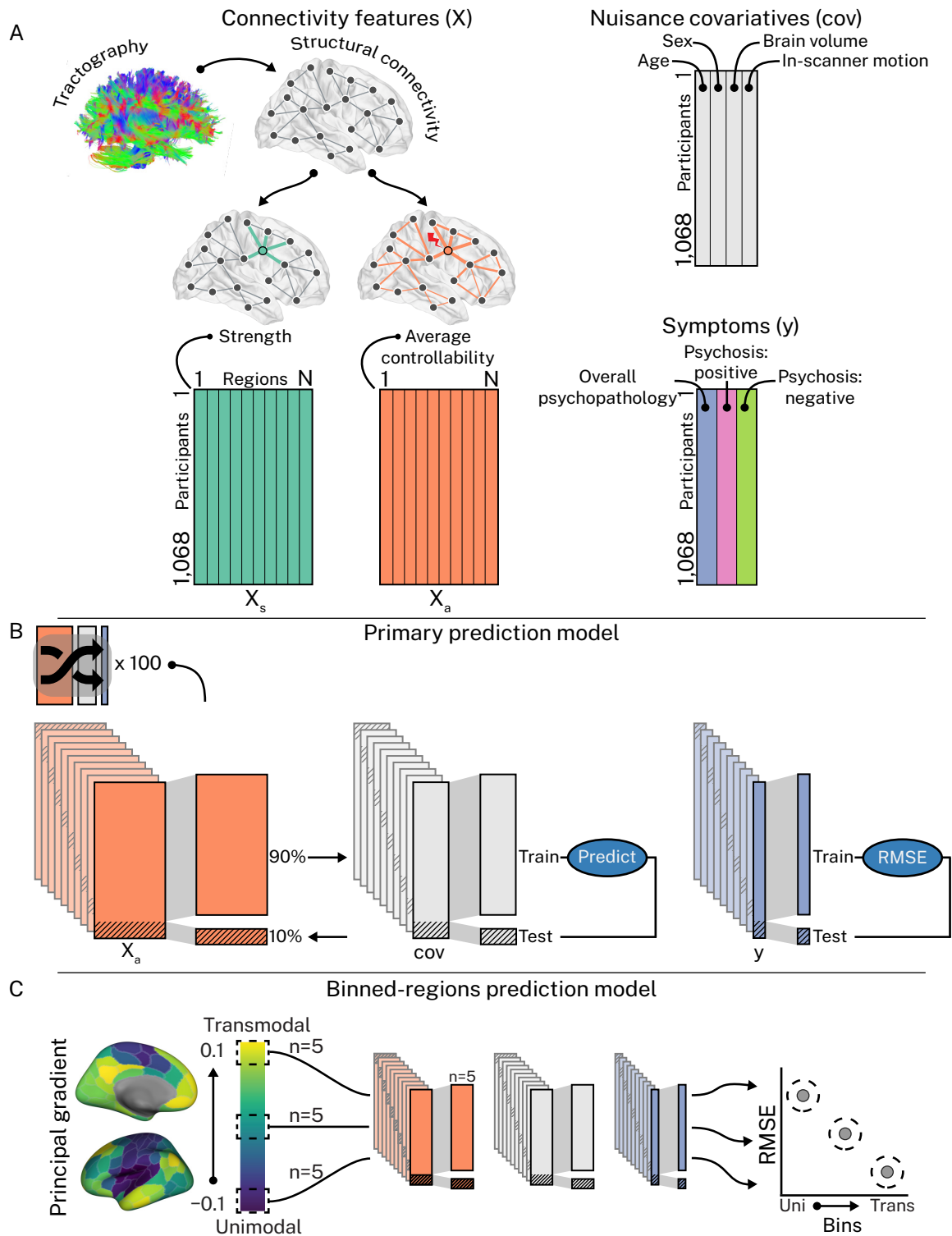

**Figure 1. Machine learning prediction models. A**, We combined regional structural connectivity features (strength, $X_s$; average controllability, $X_a$), nuisance covariates (cov; age, sex, total brain volume, and in-scanner motion), and symptom dimensions ($y$; overall psychopathology, positive psychosis spectrum symptoms, and negative psychosis spectrum symptoms) into two main prediction models. Note, we ran each combination of $X$ and $y$ separately. **B**, In our *primary prediction model*, $X$ was used to predict $y$, controlling for age, sex, brain volume, and in-scanner motion, via 100 repeats of 10-fold cross-validation, each repeat using a different random split of the data. This model provided robust estimates of prediction performance that could be compared across combinations of $X$ and $y$. **C**, In our *binned-regions prediction model*, $X$ was used to predict $y$, controlling for age, sex, brain volume, and in-scanner motion, using non-overlapping subsets of 5

regions sampled from the principal functional gradient. This model enabled examination of the regional contribution to prediction performance.

*Characterizing the unique inter-individual variation introduced to regional connectivity profiles through the examination of indirect connections*

Next, we sought to quantify the extent to which analyzing indirect structural connections to a region (average controllability) revealed unique inter-individual variation compared to analyzing only the direct connections to a region (strength). To achieve this goal, we calculated the regional cross-subject Pearson's correlation between strength and average controllability in the full sample. Then, we examined how these regional correlation values varied over the principal cortical gradient. Lower regional correlations indicated greater unique variance across strength and average controllability, suggesting greater influence of indirect connections to a region's connectivity profile.

*Varying the contribution of indirect connectivity to average controllability*

We conducted one final analysis designed to stringently examine the extent to which indirect regional connectivity influenced prediction performance for average controllability. We recalculated average controllability, increasingly restricting its access to the indirect structural connections of a region by varying the $c$ parameter in *Eq. 3* [1 (default), 10, 100, 1000, 10000]. Increasing $c$ prior to estimating average controllability causes increasingly rapid decay of all the eigenmodes of an *A* matrix (42) (see Figure S3). This decay, in turn, diminishes a region's capacity to distribute energy throughout the network, restricting its capacity to use the indirect connections to drive changes in brain state. We reasoned that this diminished access to indirect connections would reduce the predictive performance observed for average controllability. As such, we examined the impact of varying the $c$ parameter on prediction performance by repeating our binned-regions prediction model for each value of $c$.

**RESULTS**

*Participants*

Sample demographics, including counts of individuals who met diagnostic criteria for lifetime presence of a broad array of clinical symptoms (50,51), are shown in Table 1 (see also Figure S4 for mean symptom dimensions as a function of psychopathology groups).

Table 1. Summary of demographic and psychopathology data

|  | Sample ($n$ = 1,068) |
|---|---|
| Age, Year, Mean (±SD) | 15.36 (±3.42) |
| Sex, $n$, (%) |  |
| Male | 485 (45.51) |
| Female | 582 (54.49) |
| Psychopathology categories, $n$ (%) |  |
| Psychosis spectrum | 303 (28.37) |
| Manic episode | 11 (1.03) |
| Major depressive episode | 156 (14.01) |
| Bulimia | 4 (0.37) |
| Anorexia | 15 (1.40) |
| Social anxiety disorder | 261 (24.44) |
| Panic | 10 (0.94) |
| Agoraphobia | 61 (5.71) |
| Obsessive compulsive | 30 (2.81) |
| Post-traumatic stress | 136 (12.73) |
| Attention deficit hyperactivity | 168 (15.73) |
| Oppositional defiant | 353 (33.05) |
| Conduct | 85 (7.96) |

Owing to comorbidity, individual participants may be present in more than 1 category of lifetime prevalence.

*Examining indirect regional structural connectivity with network control theory enables better prediction of positive psychosis spectrum symptoms*

First, we examined how indirect connections to a region affected predictive performance for each symptom dimension by comparing RMSE for strength (direct only) and average controllability (direct and indirect). We found that strength non-significantly outperformed average controllability for overall psychopathology (Figure 2A); average controllability significantly outperformed strength for psychosis-positive scores (Figure 2B); and strength and average controllability performed similarly for psychosis-negative scores (Figure 3C). These effects were reproduced under our secondary prediction model (Figure S7). Critically, we found that only average controllability was able to predict psychosis-positive scores beyond chance levels (Figure 2B; ◉*, $p < 0.05$, FDR-corrected). Thus, in partial support of hypothesis 1, our results demonstrate that regional summaries of structural connectivity that incorporate both direct and indirect properties of connectivity were critical for predicting positive PS symptoms.

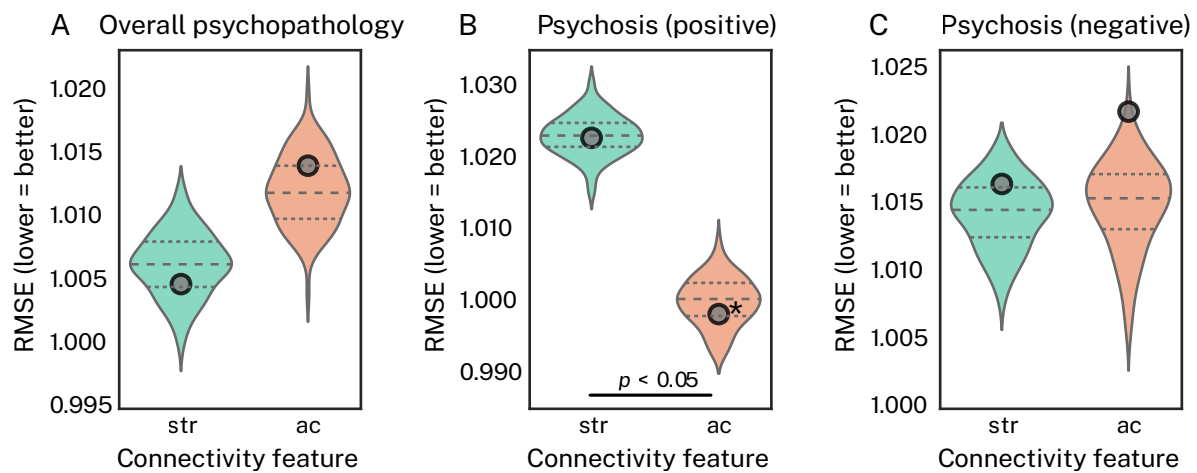

**Figure 2. Average controllability better predicts positive PS symptoms compared to strength, which better predicts overall psychopathology.** Each subplot shows prediction performance, measured via root mean squared error (RMSE; lower = better), under our *primary prediction model* using a kernel ridge regression estimator. The data shown here are the same as those presented in Figure S4, but they have been re-organized to directly compare strength (str) and average controllability (ac) for each symptom dimension via an exact test of differences. Furthermore, true RMSE from our *null prediction model* is overlaid (◉) in each subplot to illustrate which $(X, y)$ combinations yielded prediction performance above chance levels (* = $p < 0.05$, FDR-corrected). **A**, For overall psychopathology, strength outperformed average controllability, but this effect was not significant. **B**, For positive psychosis spectrum symptoms, average controllability significantly outperformed strength, and average controllability was predictive beyond chance levels. **C**, For negative psychosis spectrum symptoms, strength and average controllability performed comparably.

*Variance in indirect structural connections is increasingly relevant in transmodal association cortex*

The above results underscored the importance of incorporating indirect structural connections into the prediction of positive PS symptoms. Next, we characterized where along the cortical gradient inter-individual variation in indirect connectivity was most pronounced by correlating strength and average controllability. In support of hypothesis 2, we found that regional correlations between strength and average controllability decreased as a function of the principal gradient (Figure 3A; Pearson's *r* = -0.55). Specifically, sensorimotor regions at the unimodal end of the cortex showed the strongest correlations (regional Pearson's *r* between strength and average controllability ~ 0.7), while regions in the transmodal association cortex showed weaker correlations (regional Pearson's *r* between strength and average controllability ~ 0.5). This result offers two insights. First, while strength and average controllability are always positively correlated across the brain, those correlations are not redundant, suggesting that variance in the indirect structural connections captured by average controllability are relevant at all levels of the cortex. Second, regions in transmodal cortex have more complex profiles of indirect connectivity that drive greater divergence between average controllability and strength.

*Indirect connectivity from transmodal association cortex underpins better prediction performance of positive PS symptoms*

Having found that average controllability and strength diverged most in transmodal cortex, we tested our third hypothesis: that this divergence in transmodal cortex would drive improved performance for predicting positive PS symptoms using average controllability (see Figure S8 for psychosis-negative scores). Specifically, we examined the extent to which RMSE varied as a function of a cortical gradient separating unimodal sensorimotor cortex from transmodal association cortex. As expected, for average controllability, we found prediction was better for bins of regions sampled from transmodal cortex than for those sampled from unimodal cortex (Figure 3B; Pearson's *r* = -0.57), suggesting that regions in transmodal cortex were associated with better prediction performance. Critically, we found no such relationship for strength (Pearson's *r* = 0.07). These results were robust to bin size (see Table S3).

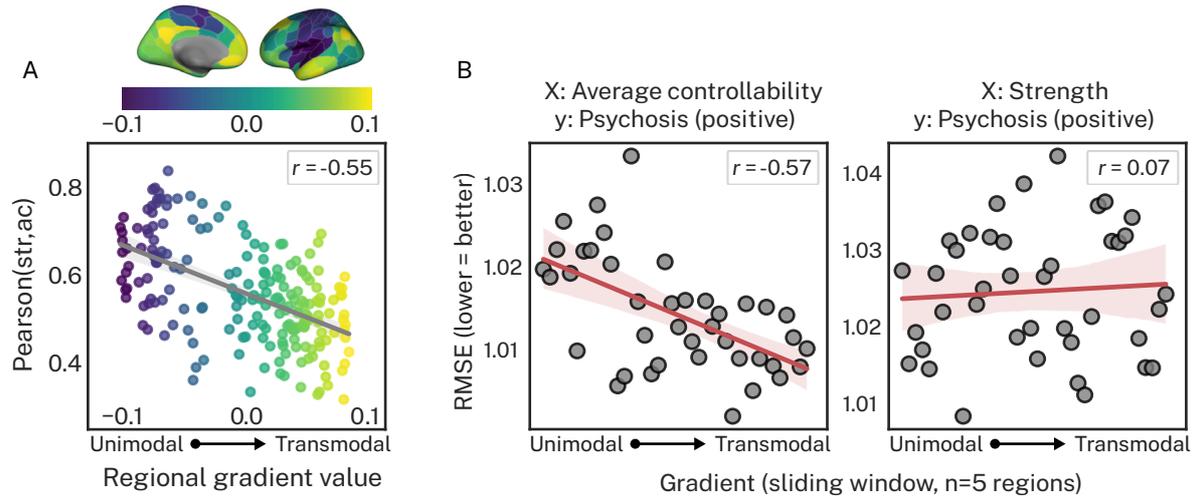

**Figure 3. Average controllability and strength are less correlated in transmodal association cortex compared to unimodal sensorimotor cortex, and prediction performance for average controllability, not strength, is better in association cortex. A**, The cross-subject correlation between strength (str) and average controllability (ac) decreases as regions traverse up the principal cortical gradient. Thus, average controllability and strength show the greatest amount of unique variance in association cortex. **B**, Performance from our *binned-regions prediction model* for average controllability (left) and strength (right) predicting positive psychosis spectrum symptoms. Prediction performance for average controllability improved as a function of the cortical gradient, whereas strength did not. Thus, the best predictive performance of positive psychosis spectrum symptoms was observed for average controllability in association cortex.

*Restricting average controllability to direct structural connection reduces prediction performance*

As a final stringent test of our hypotheses, we recalculated average controllability, increasingly constraining its access to indirect regional connections (increasing $c$ in *Eq. 3*). Figure 4A shows that increasing $c$ resulted in stronger correlations between strength and average controllability (y-axes) and diminished the spatial effect of the principal cortical gradient. Thus, not only do strength and average controllability become increasingly redundant at higher values of $c$, their unique variance becomes less differentiable as a function of the cortical hierarchy. Critically, the correlation between performance for our binned-regions prediction model and the cortical gradient also decreased at greater $c$ (Figure 4B). Together, these results provide evidence that, compared to strength, the capacity of average controllability to index direct and indirect aspects of regional structural connectivity, particularly in transmodal cortex, drives the superior predictive performance for positive PS symptoms.

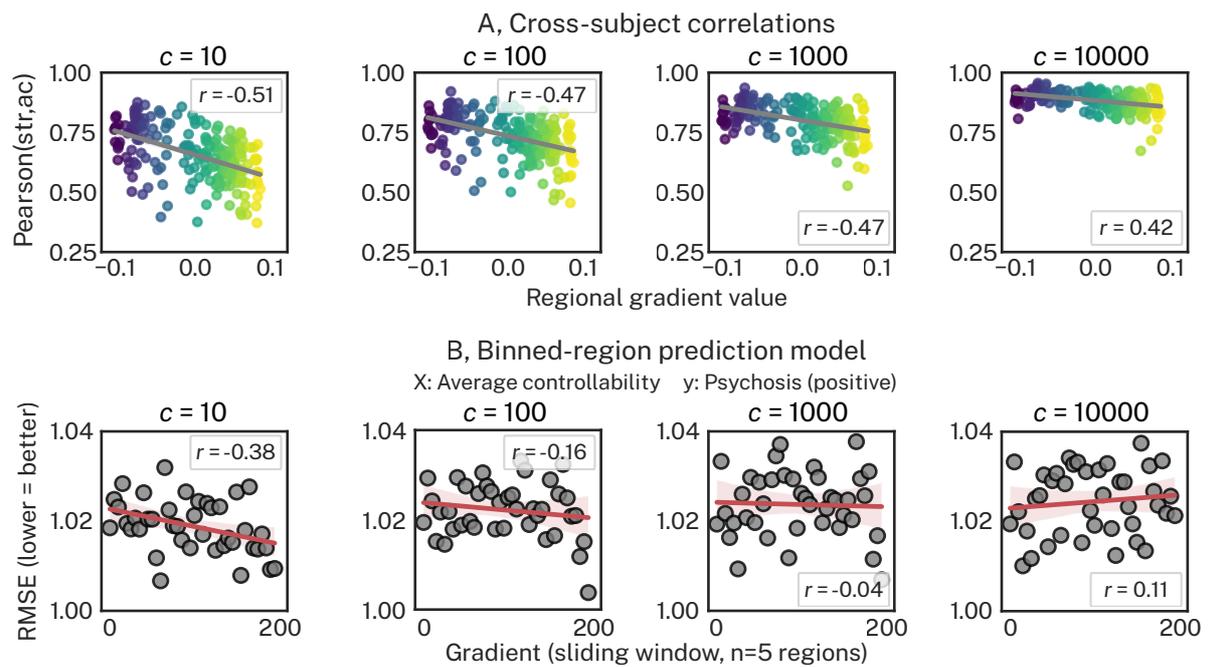

**Figure 4. Restricting average controllability to direct structural connections increases redundancy with strength and reduces prediction performance in association cortex.** Columns represent different $c$ parameters in *Eq. 3*; greater values of $c$ correspond to greater restriction of average controllability to access only direct structural connections. **A**, The cross-subject correlation between strength (str) and average controllability (ac) as a function of the principal cortical gradient. Increases in $c$ resulted in increases to the cross-subject correlations (y-axes) and a reduction in the spatial correlations with the cortical gradient, wherein correlations in the transmodal association cortex became increasingly redundant. **B**, Performance for the *binned-regions prediction model* for average controllability predicting positive psychosis spectrum symptoms. The spatial correlation between prediction performance and the cortical gradient diminished with increasing $c$, yielding lower prediction performance in association cortex.

**DISCUSSION**

A focus on examining inter-individual variation in direct regional structural connectivity has generated an incomplete picture of the pathophysiology of the PS. Here, using Network Control Theory (NCT) (9,42), we investigated the differential contributions of direct and indirect properties of regional connectivity profiles to predict positive and negative PS symptoms. We found that average controllability better predicted positive PS symptoms compared to strength, while strength and average controllability predicted negative PS symptoms to a similar degree. However, only the pairing between average controllability and positive PS symptoms yielded predictive performance better than chance levels, suggesting that robust prediction of positive PS symptoms required characterization of both direct and indirect connectivity. Additionally, in support of our hypotheses, we found that strength and average controllability exhibited the greatest amount of unique inter-individual variance in transmodal cortex, and that this unique variance linked to improved predictive performance of positive PS symptoms for average controllability. Finally, in a sensitivity analysis we systematically restricted the extent to which average controllability could access indirect connectivity. This restriction reduced both the unique covariance with strength as well as the predictive performance in transmodal cortex, bringing both more in line with that observed in unimodal cortex. Overall, our results demonstrate that NCT can quantify and probe the complex indirect connectivity pathways that stem from transmodal cortex, and that capturing this complexity can help understand positive PS symptoms in youth.

*Predicting positive psychosis symptoms using network control theory*

The structural connectivity correlates of the PS are increasingly well studied (32–39,68–70). Compared to summaries of direct connectivity, incorporating indirect connections through NCT has received relatively little attention; to our knowledge, only one previous study examined average controllability in bipolar disorder, reporting reductions compared to healthy controls (47). Here, we found that indexing indirect connections through average controllability was able to significantly predict positive PS symptoms out-of-sample where strength could not.

Furthermore, consistent with literature implicating hub dysconnectivity in schizophrenia (32,35,36), we found that average controllability showed better predictive performance in transmodal association cortex compared to unimodal sensorimotor cortex. While simple structural connectivity features like strength are readily interpretable from a network perspective, they lack an explicit model of macroscale brain function (71). Further, while other metrics apart from average controllability exist that also capture indirect connections (18,72), many of them similarly lack an explicit model of brain dynamics. By contrast, average controllability models a region's capacity to distribute input energy throughout the brain to drive changes in brain state (9), facilitating the capacity to predict the brain-wide response to external stimulation (10,73). Hence, our results demonstrate that the continued examination of NCT has potential implications for clinical treatment. For instance, external neurostimulation techniques, such as transcranial magnetic stimulation, are increasingly being investigated as treatment modalities for PS-related conditions, and candidate stimulation sites typically occupy transmodal cortex (74,75). Indeed, the analysis of neurostimulation data with NCT has begun to show promise (76).

*Predicting negative psychosis spectrum symptoms*

Unlike positive PS symptoms, predictive performance for negative PS symptoms was comparable across strength and average controllability, suggesting that capturing the indirect aspects of regional connectivity offered no substantive improvement to prediction. However, our null prediction model revealed no significant prediction effects for psychosis-negative scores, suggesting we were unable to predict negative PS symptoms with either brain feature. This failure may be due to the fact that, in our model, the psychosis-negative factor explained less variance in symptom data compared to the psychosis-positive factor (see Table S1), and hence our estimate of negative PS symptoms was perhaps noisier than our estimate of positive PS symptoms. Similar disparities in variance explained between positive and negative psychosis dimensions have also been reported in previous literature using principal component analysis (77). Thus, future work improving the modeling of variance in negative PS symptoms is needed and may yield improved predictive performance in brain-based association studies.

*Analysis of indirect connectivity is crucial in transmodal cortex*

Analysis of the principal cortical gradient (26) revealed that the cross-subject correlation between strength and average controllability was lowest in transmodal cortex. While previous research has demonstrated that high average controllability depends on high strength (9), our study is the first to illustrate that inter-individual covariance between these features has a strong spatial component. This result is consistent with the idea that brain regions' structural properties, and potential strategies for affecting change in regional activity, vary markedly over the cortical hierarchy. For instance, work in rodents illustrates that transmodal cortex broadly occupies the top-most level of the cortical hierarchy (31), wherein regions exert regulatory control over the lower levels through far-reaching and cascading sets of feedback projections. These differential roles across the mouse cortical hierarchy are also reflected by distinct microstructural properties, including variations to gene expression and cytoarchitecture (78). In human work, top-down connections from association cortex enable more efficient distribution of activity across the human brain relative to sensorimotor cortex (29). Thus, our results suggest that, compared to strength, average controllability is more sensitive to these cascading circuits of connectivity. Indeed, we found that reducing average controllability's access to indirect connections both increased the correlation with strength and reduced the spatial dependence of these correlations on the cortical gradient. Clearly, our results illustrate the value of using NCT to supplement the analysis of inter-individual differences in structural connectivity, particularly in transmodal association cortex.

Our results are also consistent with findings from work investigating relationships between structure and function of brain networks (79). At the edge-level, regional correlations between vectors of structural and functional connectivity have similarly been shown to be lower in transmodal cortex compared to unimodal cortex (80–82), and are, in general, stronger when higher-order interactions are taken into account in the structural data (29,72,79,83–85). At the regional level, strength has been shown to correlate spatially to regional frequency patterns of resting-state BOLD activity, wherein regions with higher strength have slower fluctuations compared to low-strength regions (86). Our results suggest that this spatial structure-function coupling in particular may be better elucidated through average controllability. For example, analysis of regional BOLD activity (and regional functional connectivity) is unable to isolate the influence of direct and indirect functional connections to a region; the latter are likely to be more numerous in transmodal cortex (29,31). Thus, compared to strength, the capacity to concurrently index direct and indirect structural connectivity may facilitate tighter spatial coupling to regional BOLD fluctuations (and subjects' principal gradients of functional connectivity). This improved spatial coupling, in turn, may facilitate more accurate modeling of the (a)typical changes in structure-function coupling throughout development (82).

*Limitations*

A limitation of this study is the use of a simple, noise-free, linear model of neuronal dynamics to estimate average controllability. While this assumption is an over-simplification of brain dynamics, linear models explain variance in the slow fluctuations in brain activity recorded by functional magnetic resonance imaging (87), suggesting that they approximate the kinds of data commonly used to examine brain function in psychiatry. Another potential limitation was our measurement of negative PS symptoms, which had limited construct coverage especially compared to our positive symptom measurement (see Ref. (51)), thus potentially impeding our prediction efforts. Future work could use dedicated instruments for assessing negative PS symptoms such as the Clinical Assessment Interview for Negative Symptoms (88) (CAINS).

*Conclusions*

Our results suggest that the dysconnectivity in transmodal cortex associated with positive PS symptoms reflects more than just disruptions to the direct connections among regions, and that understanding dysconnectivity along longer indirect pathways is critical to out-of-sample prediction. More broadly, our results highlight the advantages of using model-based approaches to networks such as NCT to understand dimensions of psychopathology. Continued examination of NCT and related approaches may facilitate improved predictive modeling in computational psychiatry; a goal critical to driving the field towards personalized medicine.

**CITATION DIVERSITY STATEMENT**

Recent work in several fields of science has identified a bias in citation practices such that papers from women and other minority scholars are under-cited relative to the number of such papers in the field (89–93). Here we sought to proactively consider choosing references that reflect the diversity of the field in thought, form of contribution, gender, race, ethnicity, and other factors. First, we obtained the predicted gender of the first and last author of each reference by using databases that store the probability of a first name being carried by a woman (93,94). By this measure (and excluding self-citations to the first and last authors of our current paper), our references contain 9.46% woman(first)/woman(last), 14.01% man/woman, 22.11% woman/man, and 54.42% man/man. This method is limited in that a) names, pronouns, and social media profiles used to construct the databases may not, in every case, be indicative of gender identity and b) it cannot account for intersex, non-binary, or transgender people. Second, we obtained predicted racial/ethnic category of the first and last author of each reference by databases that store the probability of a first and last name being carried by an author of color (95,96). By this measure (and excluding self-citations), our references contain 10.76% author of color (first)/author of color(last), 17.75% white author/author of color, 18.94% author of color/white author, and 52.56% white author/white author. This method is limited in that a) names and Wikipedia profiles used to make the predictions may not be indicative of racial/ethnic identity, and b) it cannot account for Indigenous and mixed-race authors, or those who may face differential biases due to the ambiguous racialization or ethnicization of their names. We look forward to future work that could help us to better understand how to support equitable practices in science.

**ACKNOWLEDGMENTS**

This study was supported by grants from the National Institute of Mental Health: R21MH106799 (D.S.B. & T.D.S.), R01MH113550 (T.D.S. & D.S.B.), and RF1MH116920 (T.D.S. & D.S.B.). Additional support was provided by R01MH120482 (T.D.S.), R01MH107703 (T.D.S.), the John D. and Catherine T. MacArthur Foundation (D.S.B.), the Army Research Office contracts W911NF-14-1-0679 and W911NF-16-1-0474 (D.S.B.), the Army Research Laboratory contract W911NF-10-2-0022 (D.S.B.), R01MH107235 (R.C.G.), R01 MH119219 (R.C.G. and R.E.G.), R01 MH119185 (D.R.R.), R01 MH120174 (D.R.R.), R01MH113565 (D.H.W.) and the Penn-CHOP Lifespan Brain Institute. The PNC was supported by RC2MH089983 and RC2MH089924. The authors acknowledge Jason Z. Kim, Jennifer Stiso, and Dr. Eli Cornblath for valuable discussions during the writing of the manuscript.

**CONFLICT OF INTEREST**

The authors declare no conflict of interest.

# Supplementary Materials for
# "Network controllability in transmodal cortex predicts psychosis spectrum symptoms"

Linden Parkes, Tyler M. Moore, Monica E. Calkins, Matthew Cieslak, David R. Roalf, Daniel H. Wolf, Ruben C. Gur, Raquel E. Gur, Theodore D. Satterthwaite, & Danielle S. Bassett

# SUPPLEMENTARY METHODS

*Participants*

The institutional review boards of both the University of Pennsylvania and the Children's Hospital of Philadelphia approved all study procedures. From the original 1,601 participants from the Philadelphia Neurodevelopmental Cohort (PNC) (1), 156 were excluded due to the presence of gross radiological abnormalities distorting brain anatomy or due to medical history that might impact brain function; those with a history of psychiatric illness were retained. An additional 365 individuals were excluded because they did not pass rigorous manual and automated quality assurance for either their T1-weighted scan or their diffusion scan (2,3). Finally, 12 and 68 participants were excluded owing to the presence of disconnected regions in their structural connectivity matrix for the 200-parcel and 400-parcel parcellations, respectively, used in the current study (see section titled *Whole brain parcellation* below). This process left final samples of 1,068 (reported in main text) and 1,032 participants. Note that these samples are larger than that commonly reported in previous studies of the PNC dataset (4–6) because, unlike previous reports, we did not exclude based on history of psychiatric illness. Indeed, previous work has illustrated that this broader coverage of the PNC yields prevalence rates of mental disorders consistent with population norms (7).

From the above samples of 1,068 and 1,032, a subsample of 926 participants was used to generate the principal functional gradient (see section titled *Principal gradient of functional connectivity* below) via resting-state functional connectivity analyses (8). These participants met quality control criteria for the resting-state functional magnetic resonance imaging (rs-fMRI) data in the PNC (see section titled *Imaging data quality control* below).

*Psychopathology dimensions*

In this study, we take a transdiagnostic dimensional approach to assessing variation in the symptoms of mental health (9–12). In particular, we extended a *p-factor* model that was previously developed based on the GOASSESS interview (13,14) and that has previously been used to study the brain (15–17). Briefly, the GOASSESS is an abbreviated and modified structured interview derived from the NIMH Genetic Epidemiology Research Branch Kiddie-SADS (18) that covers a wide variety of psychiatric symptomatology such as the occurrence of mood (major depressive episode, mania), anxiety (agoraphobia, generalized anxiety, panic, specific phobia, social phobia, separation anxiety, obsessive compulsive disorder), externalizing behavior (oppositional defiant, attention deficit/hyperactivity, conduct disorder), eating disorder (anorexia, bulimia), and suicidal thoughts and behaviors. GOASSESS was administered by trained and certified assessors. The original model used a combination of exploratory and confirmatory factor analysis to distill the 112 item-level symptoms from the GOASSESS into five orthogonal dimensions of psychopathology. The original model included a factor common to all psychiatric disorders, referred to as overall psychopathology, as well as four specific factors: anxious-misery, psychosis, externalizing behaviors, and fear.

Here, owing to emergent evidence that the positive and negative aspects of the psychosis spectrum elicit unique effects on the brain (19), we extended the above *p-factor* model in two ways. First, we included an additional five assessor-rated polytomous items (scored from 0-6, where 0 is 'absent' and 6 is 'severe and psychotic' or 'extreme' from the Scale of Prodromal Symptoms (SOPS) derived from the Structured Interview for Prodromal Syndromes (SIPS (20)) designed to measure the negative/disorganized symptoms of psychosis. These five items were (i) P5 disorganized communication, (ii) N2 avolition, (iii) N3 expression of emotion, (iv) N4 experience of emotions and self, and (v) N6 occupational functioning. Including this additional set brought the total to 117 items. Second, we split the psychosis factor into two factors, one describing the delusions and hallucinations associated with the psychosis spectrum, which we call psychosis-positive. The second psychosis factor described disorganized thought, cognitive impairments, and motivational-emotional deficits, which we call psychosis-negative for simplicity. We used confirmatory factor analysis implemented in Mplus (21) to model five specific factors of psychopathology (anxious-misery, psychosis-positive, psychosis-negative, externalizing behaviors, and fear) as well as one common factor (overall psychopathology). Note that all dimensions derived from this model are orthogonal to one another.

*Imaging data acquisition*

MRI data were acquired on a 3 Tesla Siemens Tim Trio scanner with a 32-channel head coil at the Hospital of the University of Pennsylvania. Diffusion tensor imaging (DTI) scans were acquired via a twice-refocused spin-echo (TRSE) single-shot echo-planar imaging (EPI) sequence (TR = 8100 ms, TE = 82 ms, FOV = 240mm$^2$/240mm$^2$; Matrix = RL: 128, AP: 128, Slices: 70, in-plane resolution of 1.875mm$^2$; slice thickness = 2mm, gap = 0; flip angle = 90°/180°/180°, 71 volumes, GRAPPA factor = 3, bandwidth = 2170 Hz/pixel, PE direction = AP). Our sequence utilized a four-lobed diffusion encoding gradient scheme combined with a 90-180-180 spin-echo sequence designed to minimize eddy-current artifacts (1). The sequence consisted of 64 diffusion-weighted directions with $b$ = 1000 s/mm$^2$ and 7 interspersed scans where $b$ = 0 s/mm$^2$. The imaging volume was prescribed in axial orientation and covered the entire brain.

In addition to the DTI scan, a B0 map of the main magnetic field was derived from a double-echo, gradient-recalled echo (GRE) sequence, allowing for the estimation and correction of field distortions. Prior to DTI acquisition, a 5-min magnetization-prepared, rapid acquisition gradient-echo T1-weighted (MPRAGE) image (TR = 1810ms, TE = 3.51ms, FOV = 180 x 240mm, matrix 256 x 192, voxel resolution of 1mm$^3$) was acquired for each participant.

Finally, approximately 6 minutes of rs-fMRI data was acquired using a blood oxygen level-dependent (BOLD-weighted) sequence (TR = 3000ms; TE = 32ms; FoV = 192 x 192mm; resolution 3mm isotropic; 124 volumes). These data were used solely to generate the principal cortical gradient of functional connectivity discussed in the main text (8).

*Imaging data quality control*

All DTI and T1-weighted images underwent rigorous quality control by highly trained image analysts (see Refs. (2) and (3) for details on DTI and T1-weighted imaging, respectively). Regarding the DTI acquisition, all 71 volumes were visually inspected and evaluated for the presence of artifacts. Every volume with an artifact was marked as contaminated and the fraction of contaminated volumes was taken as an index of scan quality. Scans were marked as 'poor' if more than 20% of volumes were contaminated, 'good' if more than 0% but less 20% of volumes were contaminated, and 'great' if 0% of volumes were contaminated. Regarding the T1-weighted acquisition, images with gross artifacts were considered unusable; images with some artifacts were flagged as 'decent'; and images free of artifact were marked as 'superior'. As mentioned above in the section titled *Participants*, 365 individuals were removed due to either 'poor' diffusion tensor images or 'unusable' T1-weighted images. In the main sample of 1,068, a total of 655 participants had diffusion tensor images identified as 'great', with the remaining identified as 'good', and 924 participants had T1-weighted images identified as 'superior', with the remaining identified as 'usable'. Regarding the rs-fMRI data, as in prior work (22,23), a participant's rs-fMRI run was excluded if the mean relative root mean square (RMS) framewise displacement was higher than 0.2mm, or it had more than 20 frames with motion exceeding 0.25mm.

*Structural image processing*

Structural image processing was carried out using tools included in ANTs (24). The buildtemplateparallel pipeline from ANTs (25) was used to create a study-specific T1-weighted structural template with 120 participants that were balanced on sex, race, and age. Structural images were processed in participant's native space using the following procedure: brain extraction, N4 bias field correction (26), Atropos tissue segmentation (27), and SyN diffeomorphic registration (25,28).

*Diffusion image processing*

For each participant, a binary mask was created by registering the standard fractional anisotropy mask provided by FSL (FMRIB58 FA) to the participant's mean *b*=0 reference image using FLIRT (29). To correct for eddy currents and head motion, this mask and the participant's diffusion acquisition was passed to FSL's *eddy* (30) (version 5.0.5). Diffusion gradient vectors were subsequently rotated to adjust for the motion estimated by *eddy*. Distortion correction was conducted via FSL's FUGUE (31) using the participant's field map, estimated from the *b*=0 map.

Whole-brain deterministic fiber tracking was conducted using DSI Studio (32) with a modified fiber assessment by continuous tracking (FACT) algorithm with Euler interpolation. A total of 1,000,000 streamlines were generated for each participant that were between 10mm and 400mm long. Fiber tracking was performed with an angular threshold of 45° and step size of 0.9375mm.

*rs-fMRI processing*

State-of-the-art processing of functional data is critical for valid inference (33). Thus, functional images were processed using a top-performing preprocessing pipeline implemented using the eXtensible Connectivity Pipeline (XCP) Engine (23), which includes tools from FSL (31,34) and AFNI (35). This pipeline included (1) correction for distortions induced by magnetic field inhomogeneity using FSL's FUGUE utility, (2) removal of 4 initial volumes, (3) realignment of all volumes to a selected reference volume using FSL's MCFLIRT, (4) interpolation of intensity outliers in each voxel's time series using AFNI's 3dDespike utility, (5) demeaning and removal of any linear or quadratic trends, and (6) co-registration of functional data to the high-resolution structural image using boundary-based registration. Images were de-noised using a 36-parameter confound regression model that has been shown to minimize associations with motion artifact while retaining signals of interest in distinct sub-networks (23,33,36). This model included the six framewise estimates of motion, the mean signal extracted from eroded white matter and cerebrospinal fluid compartments, the mean signal extracted from the entire brain, the derivatives of each of these nine parameters, and quadratic terms of each of the nine parameters and their derivatives. Both the BOLD-weighted time series and the artifactual model time series were temporally filtered using a first-order Butterworth filter with a passband between 0.01 and 0.08 Hz (37).

*Whole brain parcellation*

Analyses reported in the main text were conducted using 200 regions covering the cortex that were defined using functional neuroimaging data in a previous study (38), hereafter referred as the Schaefer200 parcellation. We repeated our analyses using a higher resolution version of the Schaefer parcellation that included 400 regions covering the cortex, hereafter referred as the Schaefer400 parcellation.

*Nuisance covariates*

In this study, we used age, sex (binary), total brain volume, and mean in-scanner motion as nuisance covariates. Total brain volume was generated from the T1-weighted images using ANTs. In-scanner head motion was estimated for each participant from their diffusion sequence as relative framewise displacement (2).

Specifically, rigid-body motion correction was applied to the seven high quality $b = 0$ images interspersed throughout the diffusion acquisition. Once estimated, framewise displacement was averaged across time to create a single measure for each participant.

*Machine learning prediction models*

As discussed in the main text, we generated two $1,068 \times 200$ matrices ($X$) of regional structural connectivity features: strength ($X_s$) and average controllability ($X_a$) that were used to iteratively predict symptom dimensions ($y$) in a series of prediction models. Here, we describe these models in greater detail, including our use and comparison of both linear and non-linear regression estimators ($regr$).

*Primary prediction model*

In order to elucidate whether the link between connectivity features ($X$) and symptom dimensions ($y$) was linear or non-linear, we used both linear ridge regression (RR) and kernel ridge regression (KRR) with a radial basis function (39). Regression estimators were fit using *scikit-learn* (40) with default parameters (RR: $\alpha = 1$; KRR: $\alpha = 1, \gamma = 1/200$). For each ($regr, X, y$) combination, we assessed out-of-sample prediction performance using 10-fold cross-validation scored by root mean squared error (RMSE); we also scored our models using the correlation between the true $y$ and the predicted $y$ (corr_y) and correlated these scores with RMSE to assess consistency across scoring metrics. Models were trained using all columns of $X$ (i.e., $X_s$ or $X_a$ but not both) as input features and RMSE was averaged across folds to create a single estimate of prediction performance. We included age, sex, total brain volume, and in-scanner motion as nuisance covariates (see section titled *Nuisance covariates* above). All nuisance covariates, except for sex, were normalized using an inverse normal transformation. Nuisance covariates were controlled for by regressing their effect out of $X$ before predicting $y$. Within each fold, nuisance covariates were fit to the training data and applied to the test data to prevent leakage. Next, owing to evidence that prediction performance can be biased by the arbitrariness of a single split of the data (41), we repeated 10-fold cross-validation 100 times, each time with a different random 10-fold split. Together, this process yielded a distribution of 100 mean RMSE values for each ($regr, X, y$) combination. Schematic illustration of our *primary prediction model* is presented in Figure 1B in the main text.

*Secondary prediction model*

Our primary prediction model did not perform hyper-parameter optimization, instead relying on default settings for both regression estimators ($\alpha = 1$). This decision was motivated by our desire to incorporate nuisance covariates into our prediction model while minimizing leakage; a problem that may spuriously improve prediction performance (42). While *scikit-learn* includes tools for conducting unbiased hyper-parameter optimization via nested cross-validation (41), it is not set up to do both nested cross-validation and leakage-resistant nuisance regression concurrently. As such, we specified a *secondary prediction model* that did not include nuisance regression and instead performed hyper-parameter ($\alpha$) optimization via 10-fold nested cross-validation (Figure S1). Specifically, for each fold of the test data, the remaining training data was once more subjected to 10-fold cross-validation in order to find the best performing $\alpha$ parameter (i.e., inner-loop cross-validation). The optimal $\alpha$ parameter from this inner loop was then fit to the full training data and used to predict the test data. This approach isolates the evaluation of model performance from the optimization of $\alpha$.

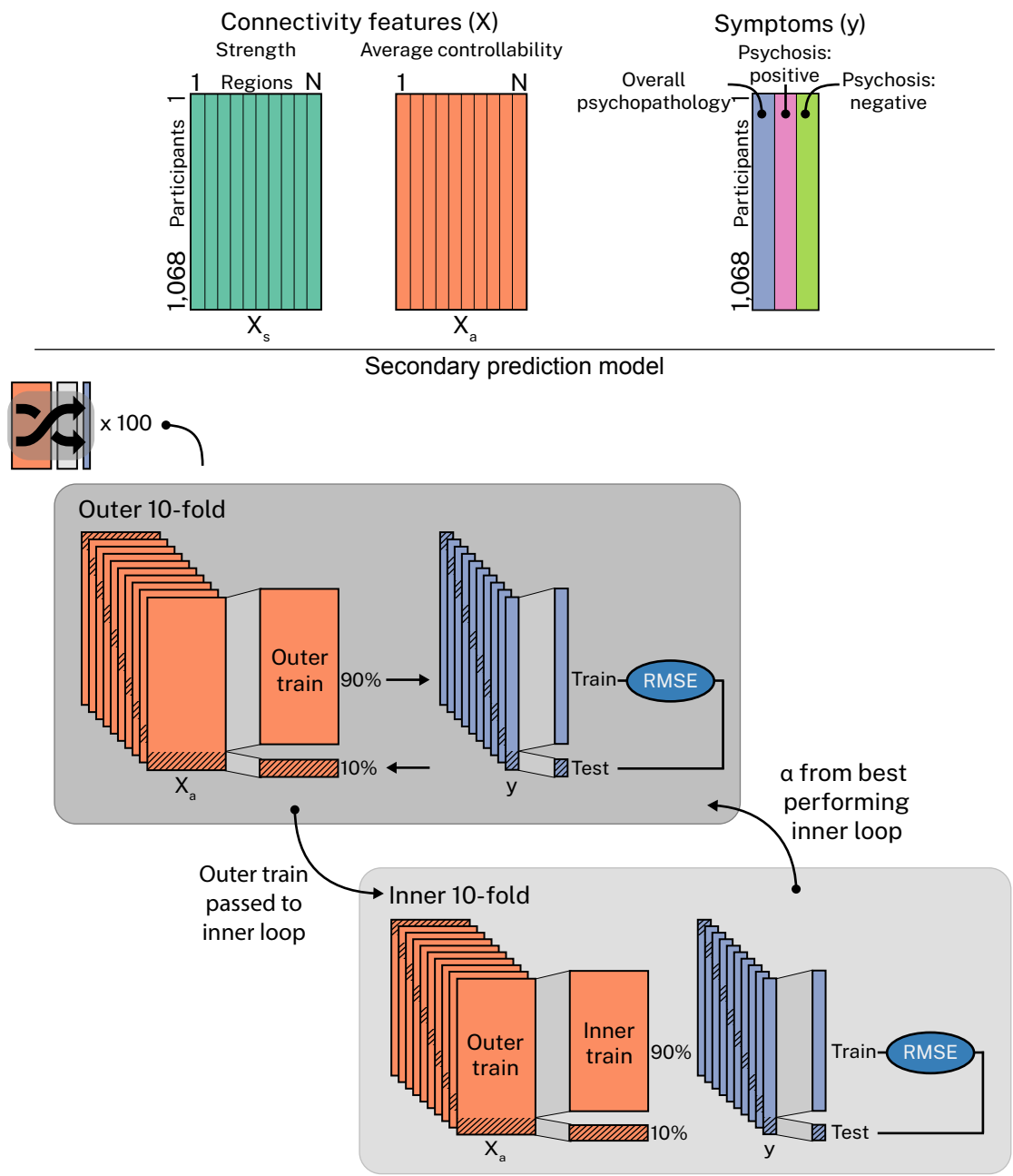

**Figure S1. Secondary prediction model.** The variable $X$ was used to predict $y$ via 100 repeats of *nested* 10-fold cross-validation; each repeat used a different random split of the data. Hyper-

parameters were optimized in the inner-loop and the best performing hyper-parameter was fed back to the outer-loop to assess performance on the test set.

*Null prediction model*

Our primary and secondary prediction models generated robust estimates of prediction performance for a given $(reg, X, y)$ combination but did not examine whether prediction performance was in itself significant. Thus, in order to test whether prediction performance was better than chance, we compared a point estimate of RMSE to the distribution of RMSE values obtained from permuted data. Specifically, we trained each $(regr, X, y)$ combination on a single cross-validation split, stratified on $y$, and then subjected the corresponding point estimate of RMSE to 5,000 random permutations, wherein $y$ was randomly shuffled. The associated *p*-values were assigned as the proportion of permuted RMSE values that were greater than or equal to our true RMSE and corrected for multiple comparisons over brain features and symptom dimensions (2*6 tests) via the Benjamini and Hochberg False Discovery Rate (FDR, *q = 0.05*) procedure (43). We refer to this model as our *null prediction model* (Figure S2).

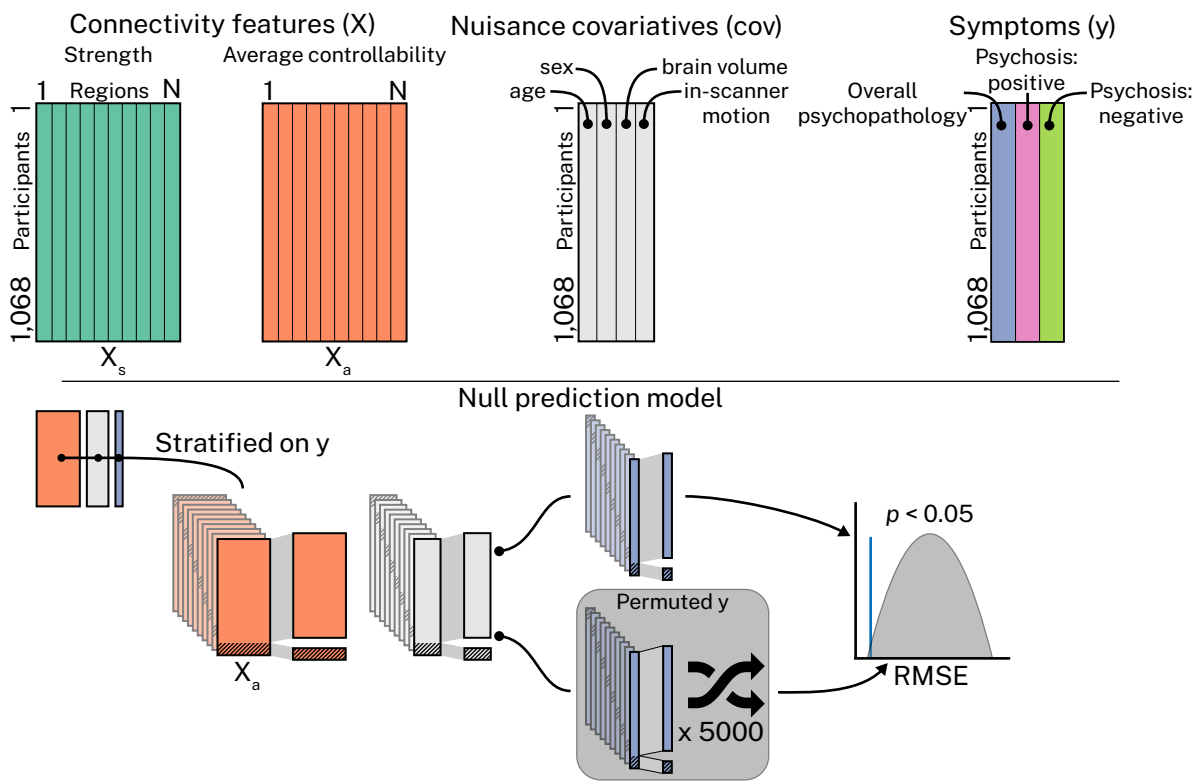

**Figure S2. Null prediction model.** The variable $X$ was used to predict $y$, controlling for age, sex, brain volume, and in-scanner motion, via a single split of the data stratified on $y$. This model was subjected to 5,000 random permutations (of $y$) to test for statistical significance.

*Principal gradient of functional connectivity*

Here, we characterized whole-brain resting-state functional connectivity according to a gradient that situates unimodal sensorimotor regions at one end and transmodal association cortex at the other (8). In particular, for each participant, processed rs-fMRI timeseries were averaged regionally for the Schaefer200 and Schaefer400 parcellations, and regional timeseries were correlated pairwise to generate functional connectomes. Correlations were estimated via Pearson's correlation coefficient and connectomes were normalized using Fisher's r-to-z transform before being averaged over participants. The principal cortical gradient was generated from this group-average connectome with the *BrainSpace* toolbox (44), using the *DiffusionMaps* approach and *normalized_angle* kernel. This process resulted in a $200 \times 1$ and a $400 \times 1$ vector for the Schaefer200 and Schaefer400 parcellation, respectively, that described each region's position along the principal gradient.

*Varying the contribution of indirect connectivity to average controllability*

In the main text, we discussed the normalization of participant's **A** matrices (*Eq. 3*) prior to estimating average controllability (*Eq. 4*). Increasing the $c$ parameter from 1 (default) to 10000 increases the rate of decay of the system, causing the system to stabilize at zero (i.e., no activity) more rapidly (45). It follows that energy distributed from a region of interest (i.e., average controllability) will be limited in its capacity to spread throughout a fast-decaying system compared to a (relatively) slow-decaying system. In turn, we reasoned that increasing $c$ would increasingly restrict average controllability's capacity to access indirect aspects of regional structural connectivity profiles, limiting the spread of activity beyond direct connections.

Here, in the Supplement, we illustrate the effect of tuning $c$ using a simple toy network. In particular, we define a toy network of 5 regions with randomly generated connection weights (Figure S3A). Then, we use this network to simulate the spread of activity from a single region of interest. We repeated this simulation to test the impact of (i) lesioning indirect connections (Figure S3B) and (ii) varying the $c$ parameter [1 (default), 10, 100, 1000, 10000] (Figure S3C). For the former, lesioning involved turning off the connection weights of edges that were not directly connected to our region of interest (i.e., the indirect connections). All code to reproduce these simulations can be found on the first author's GitHub page: https://github.com/lindenmp/linear_system_demos/blob/master/impulse_response.ipynb.

Figure S3B shows that average controllability diminished as we lesioned stronger indirect connections. Note, because all direct connections remained intact, strength is by definition unaffected. Next, Figure S3C (top) shows that activity spread the furthest throughout the toy network when $c = 1$. Moreover, average activity (over nodes) diminished to zero more rapidly over time at higher $c$ (Figure

S3C, bottom). Together, these simulations demonstrate how average controllability is sensitive to the indirect connections to a region, and that increasing $c$ limits average controllability's capacity to access these properties.

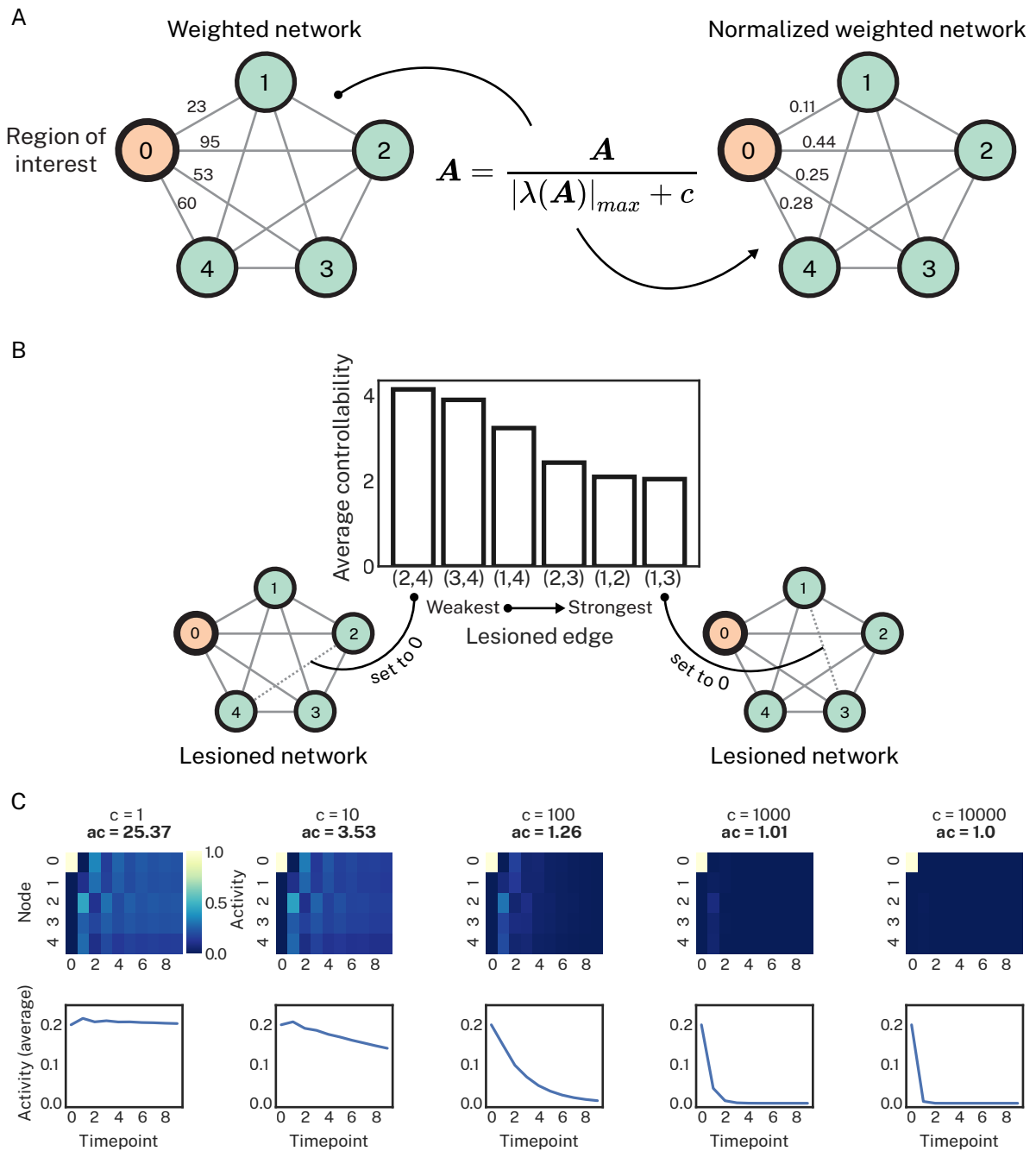

**Figure S3. Simulating the linear dynamical system and manipulating the normalization parameter ($c$). A**, A fully connected and weighted toy network comprising 5 regions was used to simulate the linear dynamical system. While all edges were weighted, only the direct connections to our region of interest are illustrated for simplicity. **B**, Average controllability was calculated for this region of interest multiple times, each time lesioning one of the indirect connections from the region of interest. Lesioning the strongest indirect connections yielded the smallest estimate of average controllability. Note that the direct connections were never lesioned in this analysis. **C**, Using the full network, we simulated the linear system over a restricted set of time steps (n=10) as a function of increasing $c$ (columns). The top row shows activity at each region at each timepoint, represented as a heatmap. The bottom row shows activity averaged over regions at each timepoint. Greater values

of $c$ caused the spread of activity to diminish to 0 increasingly quickly. This decreased activity spread corresponded to a reduction in average controllability (ac) from 25.37 to 1.0. Note that average controllability at $c = 1$ is high (25.47) owing to the size and fully connected nature of the toy network. Once a single connection is lesioned, average controllability drops dramatically.

# SUPPLEMENTARY RESULTS

*Dimensional measures of the psychosis spectrum*

Males in our sample had significantly higher psychosis-positive ($t = 2.34$, $p < 0.05$, FDR-corrected) and psychosis-negative ($t = 3.41$, $p < 0.05$, FDR-corrected) scores compared to females. Age correlated significantly with higher overall psychopathology ($r = 0.29$, $p < 0.05$, FDR-corrected) and lower psychosis-positive ($r = -0.09$, $p < 0.05$, FDR-corrected) scores. Below we illustrate model statistics (Table S1) and factor loadings (Table S2) for our bifactor model of psychopathology from which positive and negative PS symptoms were extracted.

Table S1. Factor determinacy and Omega-H scores for bifactor model of psychopathology dimensions.

| Item | General ('p') | Psychosis-positive | Psychosis-negative | Anxious-misery | Externalizing | Fear |
|---|---|---|---|---|---|---|
| Factor determinacy | 0.9927 | 0.9600 | 0.9683 | 0.9548 | 0.9661 | 0.9502 |
| OmegaH$_{subscale}$ | 0.9213 | 0.0154 | 0.0056 | 0.0004 | 0.0276 | 0.0192 |

Table S2. Factor loadings from bifactor model of psychopathology dimensions.

| | Loadings | | | | | |
|---|---|---|---|---|---|---|
| Item | General ('p') | Psychosis-positive | Psychosis-negative | Anxious-misery | Externalizing | Fear |
| psy001 | **0.657** | **0.442** | 0.000 | 0.000 | 0.000 | 0.000 |
| psy029 | **0.606** | **0.411** | 0.000 | 0.000 | 0.000 | 0.000 |
| psy050 | **0.632** | **0.220** | 0.000 | 0.000 | 0.000 | 0.000 |
| psy060 | **0.666** | **0.316** | 0.000 | 0.000 | 0.000 | 0.000 |
| psy070 | **0.637** | **0.285** | 0.000 | 0.000 | 0.000 | 0.000 |
| psy071 | **0.721** | **0.187** | 0.000 | 0.000 | 0.000 | 0.000 |
| sip003 | **0.598** | **0.522** | 0.000 | 0.000 | 0.000 | 0.000 |
| sip004 | **0.422** | **0.616** | 0.000 | 0.000 | 0.000 | 0.000 |
| sip005 | **0.593** | **0.605** | 0.000 | 0.000 | 0.000 | 0.000 |
| sip006 | **0.557** | **0.559** | 0.000 | 0.000 | 0.000 | 0.000 |
| sip007 | **0.584** | **0.608** | 0.000 | 0.000 | 0.000 | 0.000 |
| sip008 | **0.519** | **0.628** | 0.000 | 0.000 | 0.000 | 0.000 |
| sip009 | **0.615** | **0.502** | 0.000 | 0.000 | 0.000 | 0.000 |
| sip010 | **0.437** | **0.666** | 0.000 | 0.000 | 0.000 | 0.000 |
| sip011 | **0.623** | **0.607** | 0.000 | 0.000 | 0.000 | 0.000 |
| sip012 | **0.639** | **0.596** | 0.000 | 0.000 | 0.000 | 0.000 |
| sip013 | **0.605** | **0.593** | 0.000 | 0.000 | 0.000 | 0.000 |
| sip014 | **0.715** | **0.489** | 0.000 | 0.000 | 0.000 | 0.000 |
| sip027 | **0.487** | 0.000 | **0.288** | 0.000 | 0.000 | 0.000 |
| sip028 | **0.517** | 0.000 | **0.305** | 0.000 | 0.000 | 0.000 |
| sip032 | **0.758** | 0.000 | **0.188** | 0.000 | 0.000 | 0.000 |
| sip033 | **0.681** | 0.000 | **0.205** | 0.000 | 0.000 | 0.000 |
| sip038 | **0.396** | 0.000 | **0.795** | 0.000 | 0.000 | 0.000 |
| sip039 | **0.483** | 0.000 | **0.631** | 0.000 | 0.000 | 0.000 |
| SIP030 | **0.524** | 0.000 | **0.383** | 0.000 | 0.000 | 0.000 |
| SIP035 | **0.714** | 0.000 | **0.302** | 0.000 | 0.000 | 0.000 |
| SIP037 | **0.387** | 0.000 | **0.395** | 0.000 | 0.000 | 0.000 |
| SIP041 | **0.459** | 0.000 | **0.846** | 0.000 | 0.000 | 0.000 |
| SIP043 | **0.496** | 0.000 | **0.678** | 0.000 | 0.000 | 0.000 |
| SIP001 | **0.461** | 0.000 | **0.328** | 0.000 | 0.000 | 0.000 |
| add011 | **0.473** | 0.000 | 0.000 | 0.000 | **0.745** | 0.000 |
| add012 | **0.458** | 0.000 | 0.000 | 0.000 | **0.749** | 0.000 |
| add013 | **0.490** | 0.000 | 0.000 | 0.000 | **0.596** | 0.000 |
| add014 | **0.442** | 0.000 | 0.000 | 0.000 | **0.606** | 0.000 |
| add015 | **0.499** | 0.000 | 0.000 | 0.000 | **0.565** | 0.000 |
| add016 | **0.510** | 0.000 | 0.000 | 0.000 | **0.678** | 0.000 |
| add020 | **0.497** | 0.000 | 0.000 | 0.000 | **0.543** | 0.000 |
| add021 | **0.448** | 0.000 | 0.000 | 0.000 | **0.599** | 0.000 |
| add022 | **0.468** | 0.000 | 0.000 | 0.000 | **0.603** | 0.000 |
| agr001 | **0.611** | 0.000 | 0.000 | 0.000 | 0.000 | **0.474** |
| agr002 | **0.635** | 0.000 | 0.000 | 0.000 | 0.000 | **0.489** |
| agr003 | **0.651** | 0.000 | 0.000 | 0.000 | 0.000 | **0.421** |
| agr004 | **0.550** | 0.000 | 0.000 | 0.000 | 0.000 | **0.422** |
| agr005 | **0.523** | 0.000 | 0.000 | 0.000 | 0.000 | **0.469** |
| agr006 | **0.620** | 0.000 | 0.000 | 0.000 | 0.000 | **0.457** |
| agr007 | **0.621** | 0.000 | 0.000 | 0.000 | 0.000 | **0.286** |
| agr008 | **0.621** | 0.000 | 0.000 | 0.000 | 0.000 | **0.453** |
| cdd001 | **0.573** | 0.000 | 0.000 | 0.000 | **0.407** | 0.000 |
| cdd002 | **0.548** | 0.000 | 0.000 | 0.000 | **0.219** | 0.000 |
| cdd003 | **0.621** | 0.000 | 0.000 | 0.000 | **0.462** | 0.000 |
| cdd004 | **0.468** | 0.000 | 0.000 | 0.000 | **0.334** | 0.000 |
| cdd005 | **0.606** | 0.000 | 0.000 | 0.000 | **0.477** | 0.000 |
| cdd006 | **0.613** | 0.000 | 0.000 | 0.000 | **0.384** | 0.000 |
| cdd007 | **0.635** | 0.000 | 0.000 | 0.000 | **0.372** | 0.000 |
| cdd008 | **0.637** | 0.000 | 0.000 | 0.000 | **0.348** | 0.000 |
| dep001 | **0.760** | 0.000 | 0.000 | **0.220** | 0.000 | 0.000 |
| dep002 | **0.724** | 0.000 | 0.000 | **0.187** | 0.000 | 0.000 |
| dep004 | **0.791** | 0.000 | 0.000 | **0.031** | 0.000 | 0.000 |
| dep006 | **0.775** | 0.000 | 0.000 | **0.034** | 0.000 | 0.000 |
| gad001 | **0.506** | 0.000 | 0.000 | **0.377** | 0.000 | 0.000 |
| gad002 | **0.554** | 0.000 | 0.000 | **0.404** | 0.000 | 0.000 |
| man001 | **0.743** | 0.000 | 0.000 | **-0.517** | 0.000 | 0.000 |
| man002 | **0.744** | 0.000 | 0.000 | **-0.567** | 0.000 | 0.000 |
| man003 | **0.732** | 0.000 | 0.000 | **-0.523** | 0.000 | 0.000 |
| man004 | **0.771** | 0.000 | 0.000 | **-0.456** | 0.000 | 0.000 |
| man005 | **0.767** | 0.000 | 0.000 | **-0.460** | 0.000 | 0.000 |
| man006 | **0.689** | 0.000 | 0.000 | **-0.487** | 0.000 | 0.000 |
| man007 | **0.808** | 0.000 | 0.000 | **-0.241** | 0.000 | 0.000 |
| ocd001 | **0.844** | 0.000 | 0.000 | **0.197** | 0.000 | 0.000 |
| ocd002 | **0.807** | 0.000 | 0.000 | **0.125** | 0.000 | 0.000 |
| ocd003 | **0.709** | 0.000 | 0.000 | **0.209** | 0.000 | 0.000 |
| ocd004 | **0.826** | 0.000 | 0.000 | **0.060** | 0.000 | 0.000 |
| ocd005 | **0.822** | 0.000 | 0.000 | **0.115** | 0.000 | 0.000 |
| ocd006 | **0.843** | 0.000 | 0.000 | **0.107** | 0.000 | 0.000 |
| ocd007 | **0.665** | 0.000 | 0.000 | **0.143** | 0.000 | 0.000 |
| ocd008 | **0.766** | 0.000 | 0.000 | **0.131** | 0.000 | 0.000 |
| ocd011 | **0.712** | 0.000 | 0.000 | **0.196** | 0.000 | 0.000 |
| ocd012 | **0.721** | 0.000 | 0.000 | **0.134** | 0.000 | 0.000 |
| ocd013 | **0.699** | 0.000 | 0.000 | **0.119** | 0.000 | 0.000 |
| ocd014 | **0.763** | 0.000 | 0.000 | **0.061** | 0.000 | 0.000 |
| ocd015 | **0.732** | 0.000 | 0.000 | **0.092** | 0.000 | 0.000 |
| ocd016 | **0.714** | 0.000 | 0.000 | **0.150** | 0.000 | 0.000 |
| ocd017 | **0.719** | 0.000 | 0.000 | **0.090** | 0.000 | 0.000 |
| ocd018 | **0.629** | 0.000 | 0.000 | **0.095** | 0.000 | 0.000 |
| ocd019 | **0.561** | 0.000 | 0.000 | **0.073** | 0.000 | 0.000 |
| odd001 | **0.588** | 0.000 | 0.000 | 0.000 | **0.436** | 0.000 |
| odd002 | **0.573** | 0.000 | 0.000 | 0.000 | **0.515** | 0.000 |
| odd003 | **0.532** | 0.000 | 0.000 | 0.000 | **0.568** | 0.000 |
| odd005 | **0.553** | 0.000 | 0.000 | 0.000 | **0.486** | 0.000 |
| odd006 | **0.634** | 0.000 | 0.000 | 0.000 | **0.397** | 0.000 |
| pan001 | **0.621** | 0.000 | 0.000 | **0.275** | 0.000 | 0.000 |
| pan003 | **0.692** | 0.000 | 0.000 | **0.156** | 0.000 | 0.000 |
| pan004 | **0.779** | 0.000 | 0.000 | **0.159** | 0.000 | 0.000 |
| phb001 | **0.276** | 0.000 | 0.000 | 0.000 | 0.000 | **0.309** |
| phb002 | **0.340** | 0.000 | 0.000 | 0.000 | 0.000 | **0.350** |
| phb003 | **0.422** | 0.000 | 0.000 | 0.000 | 0.000 | **0.282** |
| phb004 | **0.270** | 0.000 | 0.000 | 0.000 | 0.000 | **0.355** |
| phb005 | **0.186** | 0.000 | 0.000 | 0.000 | 0.000 | **0.263** |
| phb006 | **0.456** | 0.000 | 0.000 | 0.000 | 0.000 | **0.314** |
| phb007 | **0.418** | 0.000 | 0.000 | 0.000 | 0.000 | **0.388** |
| phb008 | **0.365** | 0.000 | 0.000 | 0.000 | 0.000 | **0.199** |
| scr001 | **0.494** | 0.000 | 0.000 | 0.000 | **0.163** | 0.000 |
| scr006 | **0.487** | 0.000 | 0.000 | 0.000 | **0.357** | 0.000 |
| scr007 | **0.651** | 0.000 | 0.000 | **0.210** | 0.000 | 0.000 |
| scr008 | **0.545** | 0.000 | 0.000 | 0.000 | **0.255** | 0.000 |
| sep500 | **0.462** | 0.000 | 0.000 | 0.000 | 0.000 | **0.168** |
| sep508 | **0.413** | 0.000 | 0.000 | 0.000 | 0.000 | **0.202** |
| sep509 | **0.433** | 0.000 | 0.000 | 0.000 | 0.000 | **0.226** |
| sep510 | **0.525** | 0.000 | 0.000 | **0.085** | 0.000 | 0.000 |
| sep511 | **0.310** | 0.000 | 0.000 | 0.000 | 0.000 | **0.108** |
| soc001 | **0.444** | 0.000 | 0.000 | 0.000 | 0.000 | **0.638** |

| soc002 | **0.436** | 0.000 | 0.000 | 0.000 | 0.000 | **0.557** |
| soc003 | **0.383** | 0.000 | 0.000 | 0.000 | 0.000 | **0.708** |
| soc004 | **0.449** | 0.000 | 0.000 | 0.000 | 0.000 | **0.685** |
| soc005 | **0.486** | 0.000 | 0.000 | 0.000 | 0.000 | **0.661** |
| sui001 | **0.647** | 0.000 | 0.000 | **0.185** | 0.000 | 0.000 |
| sui002 | **0.740** | 0.000 | 0.000 | **0.260** | 0.000 | 0.000 |

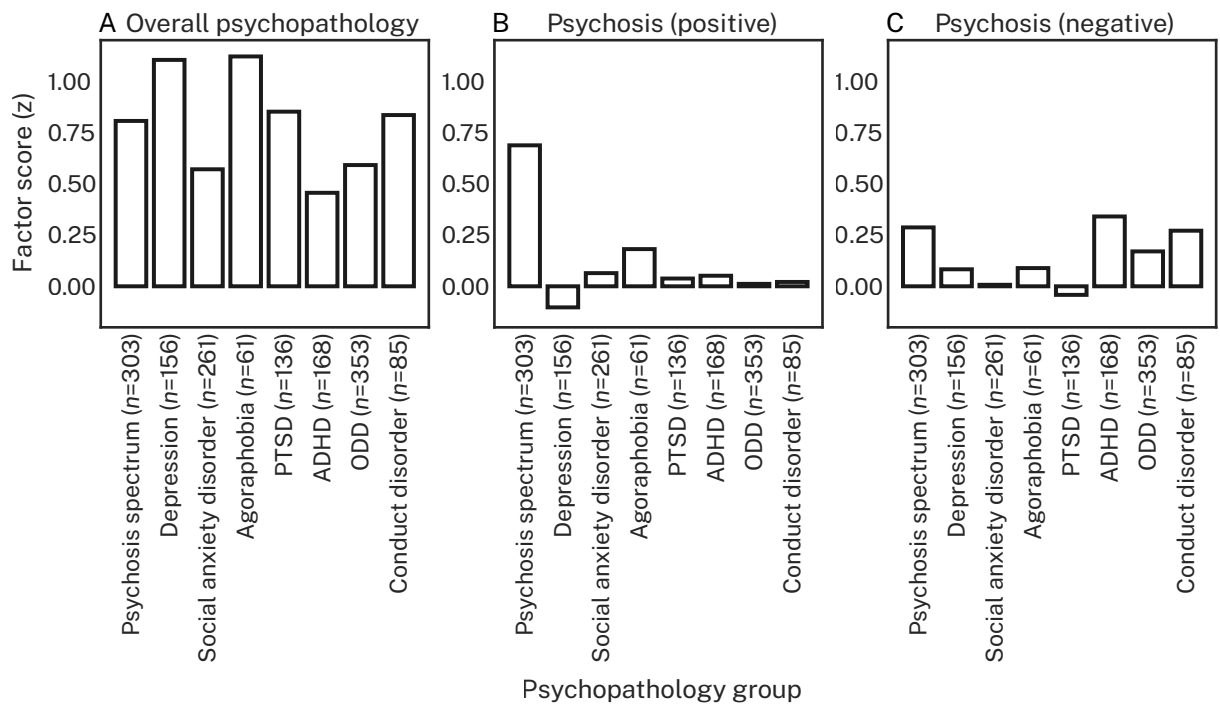

**Figure S4. Mean psychopathology symptom dimensions as a function of psychopathology groups.** Groups are the same as those presented in Table 1 in the main text. Only groups where $n \geq 50$ are shown here.

*Non-linear regression outperforms linear regression in predicting psychosis symptoms*

We sought to determine whether the mapping from structural connectivity features to symptom dimensions was best characterized by non-linear or linear regression functions. For both our *primary prediction model* (Figure S5; nuisance regression without hyper-parameter optimization), and *secondary prediction model* (Figure S6; hyper-parameter optimization without nuisance regression), kernel ridge regression (KRR) unambiguously outperformed ridge regression (RR) for all combinations of connectivity features and symptom dimensions. Given this unequivocal result, KRR was used as the sole regression estimator in all subsequent analyses. Further, we found that the correlation between RMSE and corr_y over the 100 splits of the data was $r = 0.88\pm0.03$ and $r = 0.79\pm0.10$ for the primary and secondary prediction models, respectively. These correlations were averaged over connectivity features, symptom dimensions, and regression estimators. Given these strong correlations across scoring metrics, we used RMSE as the sole estimate of prediction performance in all subsequent analyses.

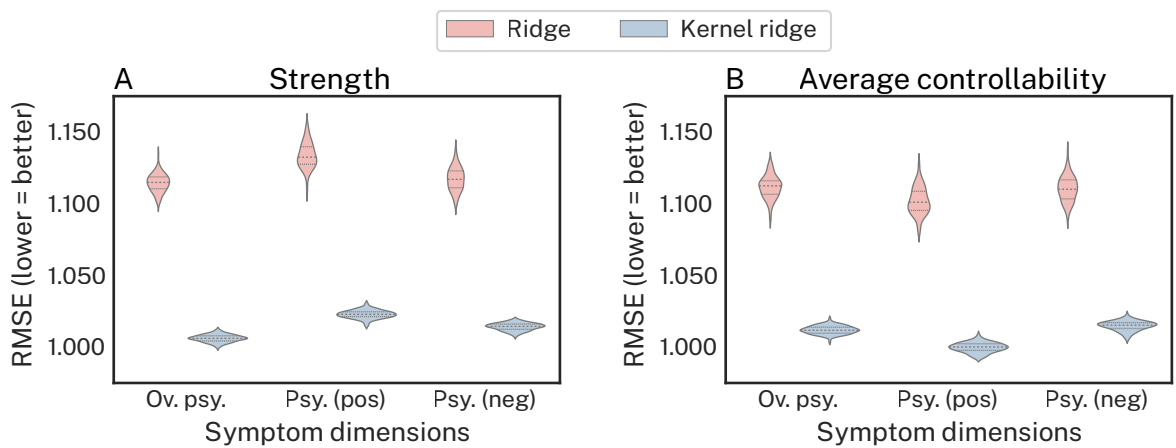

**Figure S5. Non-linear kernel ridge regression outperforms linear ridge regression at out-of-sample prediction of PS symptoms and overall psychopathology. A**, Prediction performance, measured via root mean squared error (RMSE; lower = better), for *strength* predicting each symptom dimension under our *primary prediction model*. **B**, Prediction performance for *average controllability* predicting each symptom dimension under our *primary prediction model*. For both connectivity features and all symptom dimensions, kernel ridge regression unambiguously outperformed ridge regression demonstrating that non-linear regression functions better map the relationship between regional structural connectivity and symptoms.

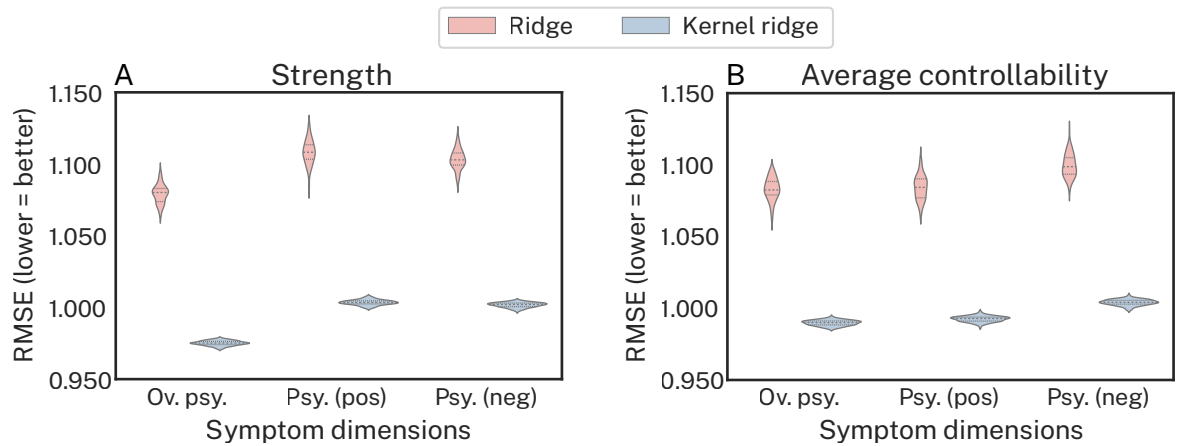

**Figure S6. Non-linear kernel ridge regression outperforms linear ridge regression at out-of-sample prediction under our *secondary prediction model*. A**, Prediction performance, measured via root mean squared error (RMSE; lower = better), for *strength* predicting each symptom dimension under our *secondary prediction model*. **B**, Prediction performance for *average controllability* predicting each symptom dimension under our *secondary prediction model*. For both connectivity features and all symptom dimensions, kernel ridge regression unambiguously outperformed ridge regression demonstrating that non-linear regression functions better map the relationship between region structural connectivity and symptoms.

*Examining indirect regional structural connectivity with network control theory enables better prediction of positive psychosis spectrum symptoms*

Similar to our primary results, Figure S7 revealed that, using KRR in our *secondary prediction model*, strength outperformed average controllability for overall psychopathology (Figure S7A), average controllability outperformed strength for psychosis-positive scores (Figure S7B), and strength and average controllability performed similarly for psychosis-negative scores (Figure S7C).

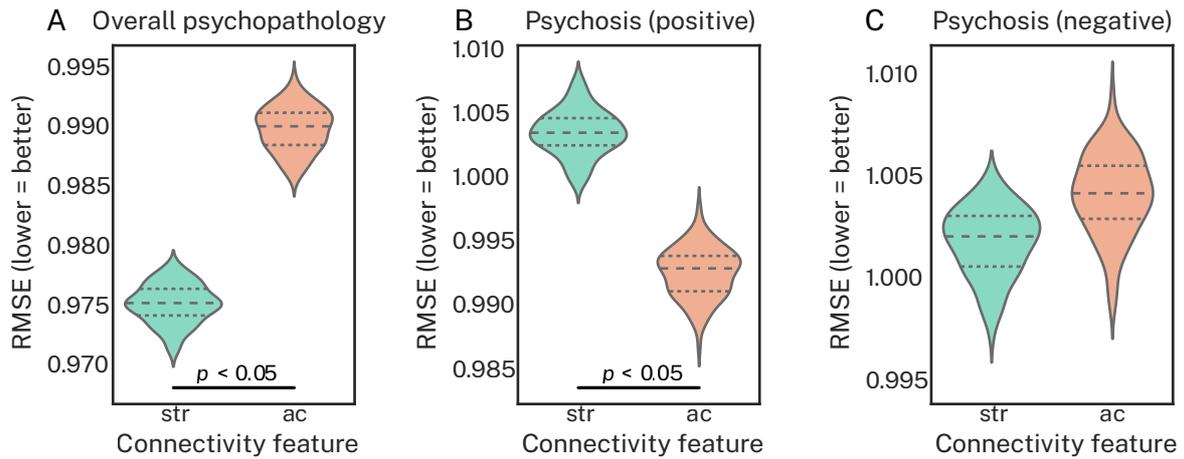

**Figure S7. Average controllability better predicts positive PS symptoms compared to strength, which better predicts overall psychopathology under our *secondary prediction model*.** Each subplot shows prediction performance, measured via root mean squared error (RMSE; lower = better), under our *secondary prediction model* using a kernel ridge regression estimator. The data shown here are the same as those presented in Figure S6, but they have been re-organized to directly compare strength (str) and average controllability (ac) for each symptom dimension via an exact test of differences. **A**, For overall psychopathology, strength outperformed average controllability. **B**, For positive psychosis symptoms, average controllability outperformed strength. **C**, For negative psychosis symptoms, strength and average controllability performed comparably.

*Indirect connectivity from transmodal association cortex underpins better prediction performance of positive PS symptoms*

In the main text we illustrated how prediction performance varied over the cortical gradient for psychosis-positive scores using our *binned-regions prediction model*. Here, we show the same analysis for the psychosis-negative dimensions (Figure S8). Broadly, we find a similar effect, wherein prediction accuracy varies over the cortical gradient for average controllability only, albeit to a lesser extent. This is consistent with the fact that we were unable to predict psychosis-negative scores significantly in our *null prediction model*. It also suggests that the relationship between connectivity features and psychosis-negative scores is, in general, weaker than that observed for psychosis-positive scores.

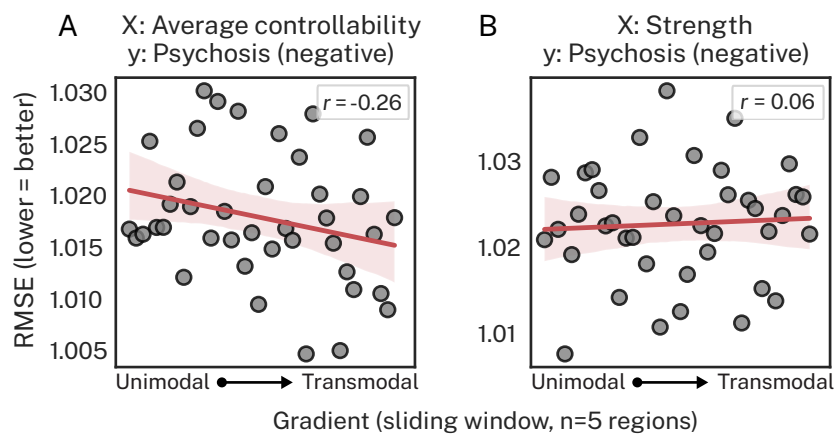

**Figure S8. Prediction performance as a function of the cortical gradient for psychosis-negative scores.** Performance from our *binned-regions prediction model* for average controllability (**A**) and strength (**B**) predicting negative psychosis symptoms.

# SENSITIVITY ANALYSES

*Binned-regions prediction model: varying bin size*

In the main text, we reported results for the *binned-regions prediction model* using bins of 5 regions sampled along the principal cortical gradient. Here, we aimed to replicate our *binned-regions prediction model* as a function of bin size. Table S3 shows that the negative correlation between bins of region and RMSE for average controllability predicting the psychosis-positive dimension was highly conserved across bin sizes.

Table S3. Pearson's correlation between out-of-sample RMSE and bins of regions sampled along the cortical gradient as a function of bin size.

| | Bin size (number of regions) | | | | | | | | | |
|---|---|---|---|---|---|---|---|---|---|---|
| | 1 | 2 | 3 | 4 | 5 | 6 | 7 | 8 | 9 | 10 |
| *Strength*, *r*-value | 0.04 | 0.08 | 0.03 | 0.00 | 0.06 | 0.07 | -0.22 | 0.03 | 0.13 | 0.16 |
| *Average Controllability*, *r*-value | -0.12 | -0.32 | -0.37 | -0.44 | -0.57 | -0.48 | -0.46 | -0.41 | -0.40 | -0.79 |

*Parcellation resolution*

In the main text, we reported results for the Schaefer200 parcellation. Here, we aimed to replicate our key results when using the Schaefer400 parcellation. Most pertinently, we assessed the extent to which our *null prediction model*, which assessed the extent to which we could predict symptom dimensions beyond chance levels, was able to reproduce the significant predictive effect between average controllability and psychosis-positive scores. We found that this effect was no longer significant under the Schaefer400 parcellation (RMSE = 1.01, $p$ = 0.36). Next, we assessed whether our *binned-regions prediction model*, which assessed the extent to which predictive performance varied over the principal cortical gradient, was able to reproduce the negative correlation between the cortical gradient and RMSE for psychosis-positive scores. We found that the correlation between bins of 5 regions and RMSE was $r$ = -0.21 for average controllability and $r$ = 0.00 for strength. Under variations of the $c$ parameter, we found that this correlation for average controllability dropped to $r$ = -0.12 and $r$ = -0.10 at $c$ = 10 and $c$ = 10000, respectively. Despite weaker effects across these results, particularly with regard to predicting psychosis-positive scores above chance levels, the general trend of average controllability, but not strength, being differentially sensitive to the cortical gradient was upheld. Similarly, the capacity to manipulate this effect of the cortical gradient on RMSE through varying $c$, was also consistent with our main results. Nevertheless, taken together, these results suggest that our results were weaker for the Schaefer400 compared to the Schaefer200 parcellation. This finding may reflect a lack of statistical power for the

Schaefer400 parcellation relative to the Schaefer200 parcellation or increased sparsity in the Schaefer400 parcellation leading to truncated indirect pathways. In any case, our results demonstrate that out-of-sample prediction of the symptoms of the psychosis spectrum may require the use of a relatively course parcellation of the brain.